# Effective Temperature in an Interacting, Externally Driven, Vertex System: Theory and Experiment on Artificial Spin Ice.

Cristiano Nisoli[1], Jie Li[2], Xianglin Ke[2], D. Garand[2], Peter Schiffer[2], Vincent H. Crespi[2]

[1]Theoretical Division and Center for Nonlinear Studies, Los Alamos National Laboratory, Los Alamos, NM, 87545 USA

[2]Department of Physics and Materials Research Institute, 104 Davey Lab, Pennsylvania State University, University Park, PA, 16802 USA

Frustrated arrays of interacting single-domain nanomagnets provide important model systems for statistical mechanics, because they map closely onto well-studied vertex models and are amenable to direct imaging and custom engineering. Although these systems are manifestly athermal, we demonstrate that the statistical properties of both hexagonal and square lattices can be described by an effective temperature based on the magnetostatic energy of the arrays. This temperature has predictive power for the moment configurations and is intimately related to how the moments are driven by an oscillating external field.

The physical meaning of temperature can be approached on several distinct levels. Macroscopically we have the intuitive notion of thermal equilibrium. Formally, temperature measures the variation of energy with disorder. In statistical mechanics temperature emerges as the Lagrange multiplier in an energy-constrained probability distribution of accessible microstates. Traditionally, a "thermal" system at equilibrium supports all three of these notions. Yet the concept of temperature has been extended to athermal systems, such as driven granular materials where the characteristic interaction energies greatly exceed the standard thermodynamic temperature [1-3] (often in the context of glassy transitions, jamming and rheology [4-7]) on the grounds that the large number of grains would still warrant statistical descriptions [8]. Effective temperatures have thus been introduced theoretically [9] and extracted from simulations of slowly sheared granular matter [10]. Recently, experimental observations of Brownian motion in vibro-fluidized grains have allowed a successful extraction of an effective temperature that increases monotonically with the magnitude of the vibratory external drive [11].

These seminal results reflect the granular kinetics of systems interacting only through hard-core repulsion and possibly friction, rather than the more complex interactions characteristic of microscopic systems and the models of statistical mechanics that describe them. In this Letter, we demonstrate that an effective temperature can be defined, extracted, used to make predictions, and related to external drive for a recently introduced nanometer-scale meta-material, 'artificial spin ice' [12-16]. In artificial spin ice, single-domain magnetic islands interact as Ising-like spins, thus allowing an *energy-based* definition of temperature in a system that can be engineered to replicate the celebrated vertex models of statistical mechanics [17-19] and whose microstates can be directly imaged.

Our artificial spin ice system is a two-dimensional array of elongated single-domain permalloy islands (80 x 220 x 25 nm, with a magnetic moment $\sim 10^7$ $\mu_B$) whose shape anisotropy defines Ising-like spins arranged along the sides of a regular lattice. The microstate of island moments in this system can be directly imaged via magnetic force microscopy (MFM) [12, 20], in contrast to the naturally occurring magnetically frustrated materials such as the spin ice pyrochlores [21-24]. The island moment configuration is not in thermal equilibrium with the surroundings, since the magnetostatic interaction and anisotropy energies of the islands are $\sim 10^5$ K, and thus thermal excitations cannot induce spin flips. However, as for granular systems, the large number of islands suggests the viability of statistical treatments, if fluctuations can be activated by an external drive. Artificial spin ice can be driven into a low-energy, interaction-dominated state [12, 20, 25-26] by rotating the sample in a decreasing magnetic field. In our experiments, the field decreases from 2000 Oe (far above coercive field) to 0 in steps of $H_s$, holding each step for 5 seconds while the sample rotates at 1000 rpm, with a reversal of field direction at each step. This demagnetization protocol was implemented for a range of field step sizes ($H_s$) and for two lattice geometries, square ice and hexagonal ice, as depicted in Fig. 1. We studied arrays with a range of different lattice constants, and each data point in figures 2-4 represents three or more MFM images, each spanning several hundred islands.

The dominant interactions in the lattice occur between neighboring islands across a given vertex. Hence we follow a previously established approach, [12, 26-27] and describe the data within a vertex model [19] wherein the system is described in terms of populations of distinct vertex types, each with a given magnetostatic energy. Square ice has 4 topologically distinct vertex types, which we call Type I, II, III, IV, with multiplicities $q_I = 2$, $q_{II} = 4$, $q_{III} = 8$, $q_{IV} = 2$ as defined in Figure 1. We call the magnetostatic self-energies of these vertices $E_I$, $E_{II}$, $E_{III}$, $E_{IV}$ with fractional populations $n_I$, $n_{II}$, $n_{III}$, $n_{IV}$; these can be extracted from MFM images. The specific vertex energy is then simply $\overline{E} = E_I n_I + E_{II} n_{II} + E_{III} n_{III} + E_{IV} n_{IV}$. Hexagonal ice has just two vertex topologies of multiplicity $q_I = 6$, $q_{II} = 2$ with specific vertex energy $\overline{E} = E_I n_I + E_{II} n_{II}$. The two lattice geometries have very different entropy vs. energy relations within the vertex model: square ice has a two-fold ground state of anti-ferromagnetically tiled Type-I vertices, whereas hexagonal ice has an extensively degenerate ground-state tiling of Type-I vertices with a substantial residual entropy. Perhaps not surprisingly, these two lattices behave differently under AC demagnetization: square ice never finds (or closely approaches) the ground state, whereas demagnetized hexagonal ice returns the vertex-model ground state with at most sparse type-II excitations.

We first consider the case of the square ice arrays, where we have studied lattices with lattice constant $a$ = 400, 440, 480, 560, 680 and 880 nm. The external field in our rotational demagnetization is initially strong enough to coerce every island into following the external field. As the magnitude of the external field is decreased, successive islands presumably begin to "fall away" from the external field, locked in by favorable magnetostatic interactions with their neighbors. Although not ergodic, the accumulation of these distinct "defects" carved in the initial uniform set of aligned type-II vertices generates a well-defined statistical system. In an isotropic, vertex-gas approximation, where each vertex is treated as an independent entity, there are $M = \dfrac{N!}{(N-D)! \prod_{i=1\ldots16} N_i!}$

ways to choose $D$ defected vertices among the $N$ vertices of a given lattice, each allocated among the sixteen vertex types $N_i$. We assume that vertices of equal topology $\alpha = I,\ldots,IV$ will be isotropically distributed according to multiplicity $q_\alpha$. We call $\rho = D/N$ and $\upsilon_\alpha = N_\alpha /D$, and maximize $S = \ln M / N$ under a vertex-energy constraint on the ensemble of defected vertices, or $\rho\sigma - \rho\ln\rho - (1-\rho)\ln(1-\rho) - \rho\beta_e(\sum_{\alpha=I}^{IV} E_\alpha \upsilon_\alpha - E)$ where $\sigma = -\sum_{\alpha=I}^{IV} \upsilon_\alpha \ln\dfrac{\upsilon_\alpha}{q_\alpha}$ is the "entropy" of the defected ensemble. We obtain a canonical distribution for the defects

$$\upsilon_\alpha = \frac{q_\alpha \exp(-\beta_e E_\alpha)}{Z(\beta_e)}, \qquad (1)$$

($Z(\beta_e)$ is defined by normalization of $\upsilon_\alpha$) as well as an expression for the auxiliary quantity $\rho$

$$\rho(\beta_e) = \frac{1}{\exp(-\sigma(\beta_e)) + 1}. \qquad (2)$$

Where $\sigma(\beta_e)$ is obtained by substitution of Eq 1 into the expression for $\sigma$. Eqs 1, 2 provide the relative vertex population densities as

$$\begin{aligned} n_I &= \rho\upsilon_I, n_{III} = \rho\upsilon_{III}, n_{IV} = \rho\upsilon_{IV} \\ n_{II} &= (1-\rho) + \rho\upsilon_{II}. \end{aligned} \qquad (3)$$

We compute the vertex energies using a "dumbbell" model (as in ref [28]), in which the magnetic dipole is treated as a finite-size dumbbell of monopoles, and we consider only interactions between monopoles converging in each vertex: energies then scale as $(a-l)^{-1}$ (where $a$ is the lattice constant and $l$ the length of the islands). By imposing $E_I = 0, E_{III} = 1$ one finds $E_{II} = (\sqrt{2}-1)/(\sqrt{2}-1/2)$ and $E_{IV} = 4\sqrt{2}/(2\sqrt{2}-1)$. In this simple dumbbell model of the energetics, the ratios between different vertex energies are independent of the array lattice constant.

As a simple test of the basic assumptions of the model above [1], we consider the quantities $\ln(5n_I/2n_{II})$ and $\ln(8n_I/2n_{III})$ as deduced from the measured $n_I$, $n_{II}$, and $n_{III}$. These quantities should be proportional to the reciprocal effective temperatures $E_{II}\beta_e$ and $E_{III}\beta_e$, since our predictions for the vertex populations (Eq. 3) at high temperatures are well approximated by a purely canonical distribution that assigns an anomalous multiplicity of 5 rather than 4 to Type-II vertices--a fact that can be checked by direct calculation but which also seems reasonable as there are 4 different multiplicities in the defected sample, and one in the background. In figure 2, we plot these two quantities against each other. A linear fit returns $E_{II}/E_{III} = 0.441$, very close to the expected theoretical value $E_{II}/E_{III} = (\sqrt{2}-1)/(\sqrt{2}-1/2) = 0.453$ obtained from the dumbbell approximation.

In Figure 3a, we plot the experimentally observed populations of each vertex type *vs.* the effective reciprocal temperature extracted from the fraction of type III vertices via $\beta_e E_{III} = \ln\dfrac{4n_I}{n_{III}}$. The figure includes data from square arrays with all of our different lattice constants and annealed at all of our different step sizes. For comparison, we also show theoretical curves for the vertex populations as a function of the effective temperatures, based on Eq. 1-3. The excellent agreement between theory and the experimental data demonstrates that the derived effective temperature has good predictive power, despite the crudity of the vertex gas approximation.

Is the effective temperature derived above only a mathematical artifice (*i.e.*, just a Lagrange multiplier within maximum likelihood), or does it provide physical information about the environment within which the system resides (i.e. the "fluidizing" external magnetic drive), as an actual physical temperature provides information about the surrounding thermodynamic bath? We found that the effective temperature of the square ice arrays can be controlled via the external drive in a way strikingly analogous to that reported for vibro-fluidized granular materials [11] – but here in a non-hard-core system with an explicit *energetic* description of interactions. As seen in Figure 3b, we find a strikingly linear dependence of $\langle \beta_e \rangle$ in the magnetic step-size of the AC demagnetization, indicating that the effective temperature description does indeed have a physical basis akin to a microscopic temperature.

We now consider the effective temperature of the hexagonal ice arrays, in which AC demagnetization consistently returns the vertex ground state (all type I vertices) for arrays of small lattice constant. For $a$ = 225, 260, 320, 425 nm, the frequency of excitations is $\sim 10^{-3}$, below experimental error. Hence hexagonal ice is a good candidate to study effective temperature only for larger lattice constants, $a$ = 650, 910, 1135, 1395, 1620 nm, wherein the occurrence of excitations $n_{II}$ is significant. As the density of excitations $n_{II}$ completely defines the thermodynamics, the introduction of an effective temperature as for the square ice, $\beta_e E_{II} = \ln (n_I/3n_{II})$ might seem only a re-parametrization with little predictive power. In Figure 4a, however, we extract $\ln(n_{II}/n_I)$ from arrays of different lattice constant $a$, but annealed with the same magnetic step $H_s$, and plot that ratio against the respective energy $E_{II}$. Somewhat surprisingly, we find a linear behavior, which suggests an effective temperature that is independent of the lattice constant. In this calculation, the vertex energies are obtained via micromagnetic calculations that describe the full vertex interaction of dipole islands [29], since we now study much larger lattices for which the dumbbell approximation (which treats only the monopole tips that converge at a vertex) is less accurate. The intercept of the fits in Figure 4a is surprisingly close to the expected $\ln(q_I / q_{II}) = \ln 3$, lending further credence to the analysis. The extracted effective temperature, $\beta_e$, is plotted in Figure 4b against the magnetic step-size $H_s$. As in the case of the square ice, we again find a remarkable linear dependence of $\beta_e$ on the anneal step size $H_s$, although with different parameters (different geometries apparently experience different effective temperatures under the same magnetic drive). These results support the physical nature of the effective temperature, one that can be generalized to multiple geometries of artificial ice systems, although the reason for the linearity in $H_s$ is not obvious.

In conclusion, we have introduced a predictive notion of effective temperature in a complex interacting system of magnetostatically interacting nanomagnets. We have found that the external drive, in the form of an agitating magnetic field, behaves as a thermal bath and controls the temperature. The formalism successfully predicts microstates on a wide spectrum of different energies and vertex populations. Unlike granular materials in which effective temperature have been previously explored, the nanomagnet arrays can be engineered to reproduce known models of statistical mechanics, and the interactions can be controlled by design, suggesting that a range of other statistical physics make be accessed in these systems.

under Contract No. DE-AC52-06NA25396 and also supported by the Army Research Office and the National Science Foundation MRSEC program (DMR-0820404) and the National Nanotechnology Infrastructure Network. We are grateful to Chris Leighton for the film deposition.

**FIGURE CAPTIONS**

Figure 1: Square and hexagonal artificial spin ice. (a) Schematics (top-left) and MFM (top-right) of the square arrays, along with the 16 vertices of the square artificial ice (bottom). (b) Schematics (top-left) and MFM (top-right) of the hexagonal arrays with the 8 vertices of the hexagonal. The white arrows on the schematics show the vertex ground states of the two lattices, and the percentages indicate the vertex multiplicity for a random moment configuration.

Figure 2: The effective temperature of the square arrays, plotted as $\ln(5n_I/2n_{II})$ vs. $\ln(8n_I/2n_{III})$: the linear fit returns a ratio very close to the theoretical value. The values of $n_I$, $n_{II}$ and $n_{III}$ are all average values obtained from the MFM images taken on the same array and at same step size. Negative temperatures (here and in Fig. 3) are in theory possible, since there exist high-energy, low-entropy states in these lattices.

Figure 3: (a) Vertex frequency from square arrays of different lattice constant and obtained with different $H_s$ plotted against their effective reciprocal temperature $\beta_e$ in units of $E_{III}$. The data are obtained by averaging the results of at least three MFM images from the same array with the same demagnetization step size. The error bars show the variation within the same array. The lines are theoretical curves from Eq. 1-3. (b) Linear dependence of $\beta_e$ as a function of the magnetic step size (data are averaged over the lattice constant $a$).

Figure 4: (a) Linear fits of $\ln(n_{II}/n_{I})$ for the hexagonal arrays *vs.* the energy $E_{II}$, for the larger lattice spacings discussed in the text. The intercept falls very close to the expected multiplicity $q_{I}/q_{II} = 3$ (*i.e.* $\ln(3)$, since the graph plots the logarithm). (b) Linear dependence of $\beta_{e}$ as a function of the magnetic step size; $\beta_{e}$ was obtained from the fitting slope in (a).

Figure 1

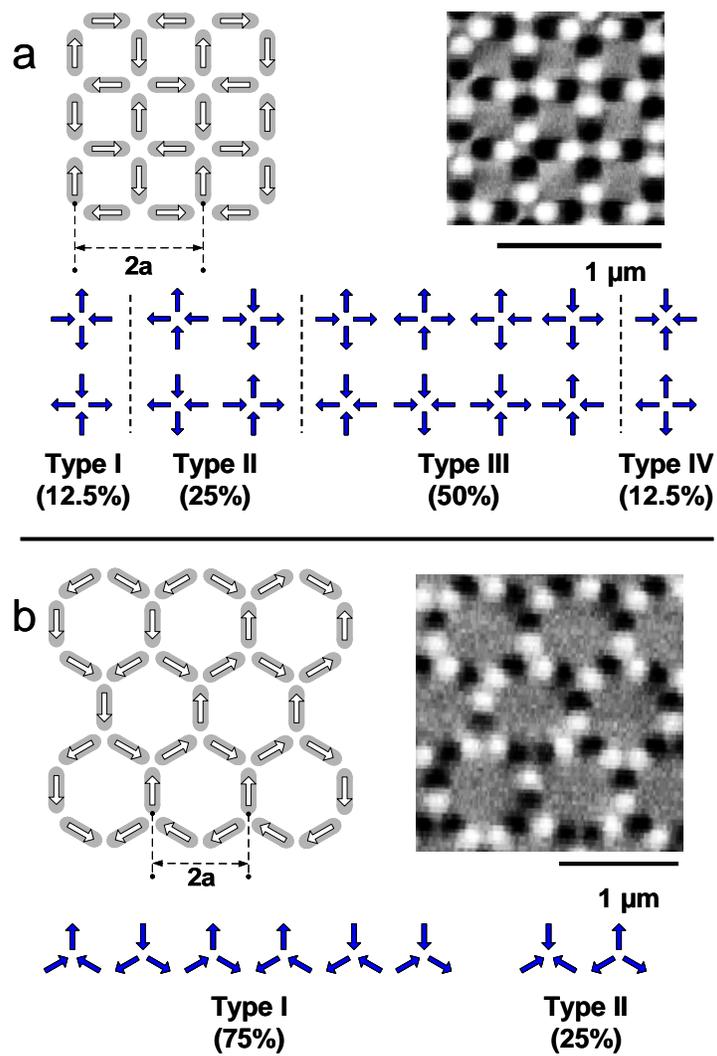

a

2a

1 μm

Type I          Type II          Type III          Type IV
(12.5%)        (25%)            (50%)              (12.5%)

b

2a

1 μm

Type I                    Type II
(75%)                    (25%)

Figure 2

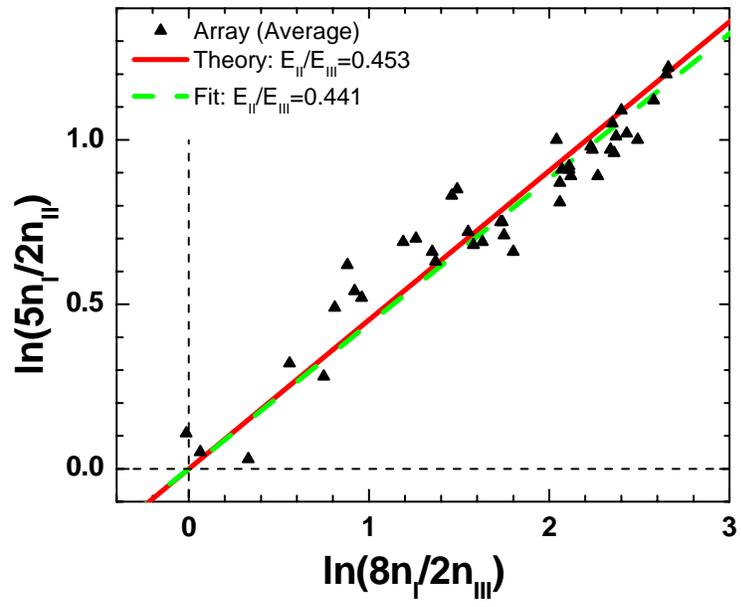

Figure 3

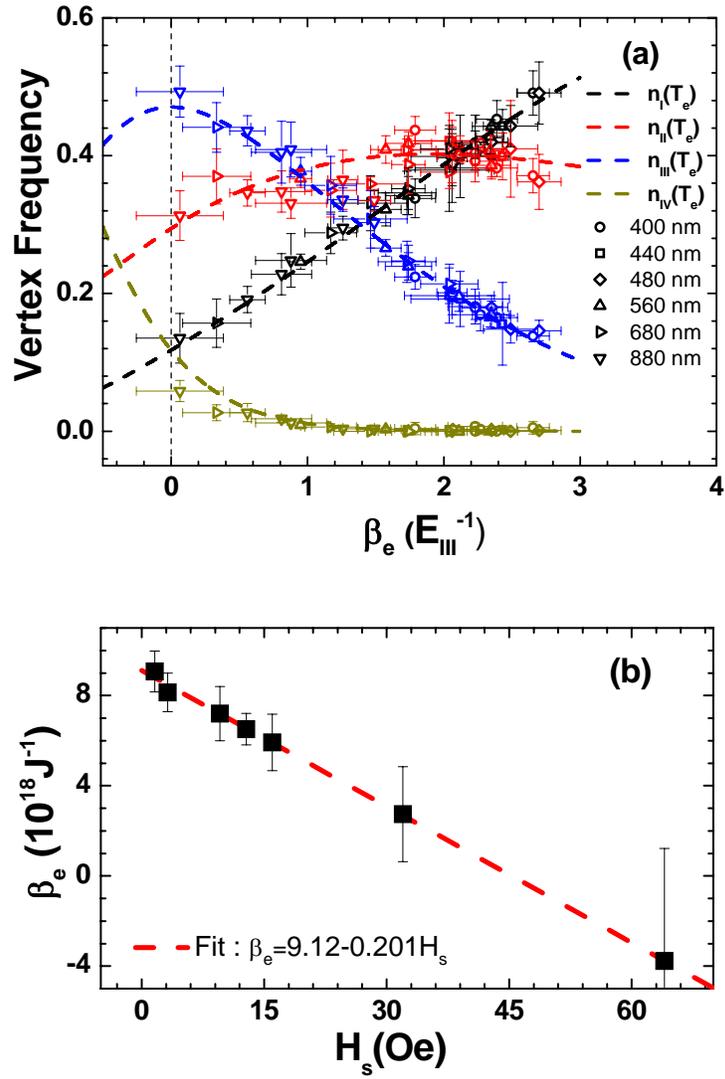

Figure 4

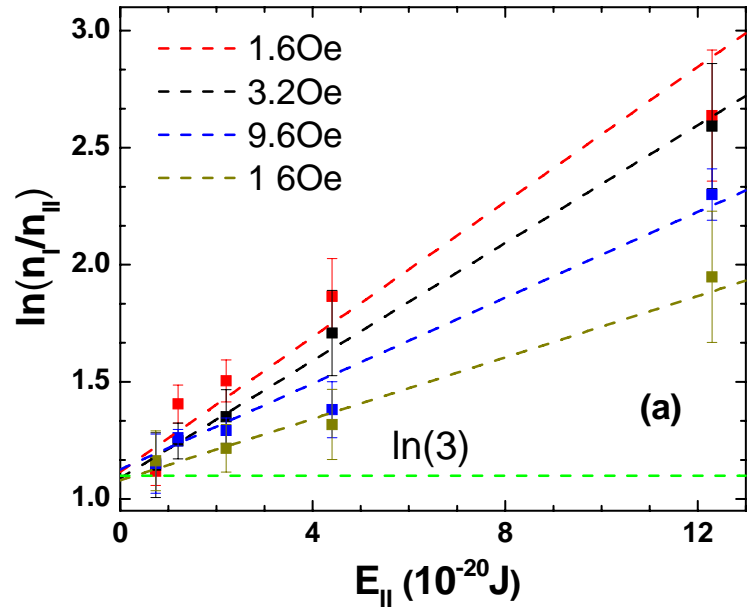

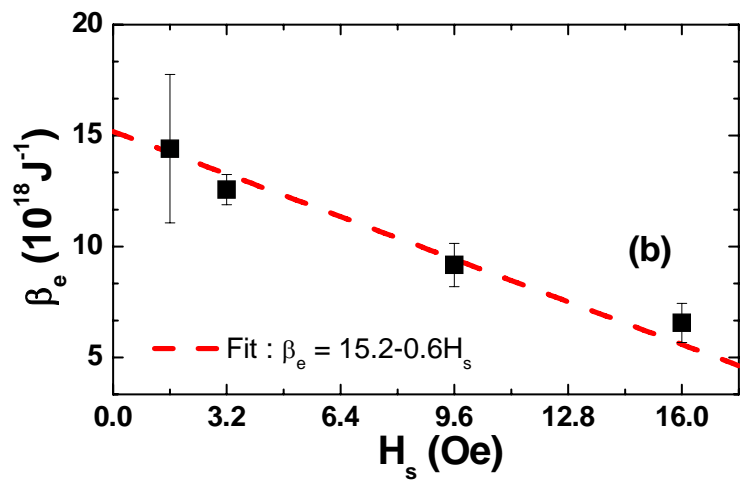